\documentstyle[preprint,aps,epsf,epsfig]{revtex}
\begin{document}
\title{Bifurcation structure of two Coupled Periodically driven
double-well Duffing Oscillators}
\author{Anatole Kenfack}
\address{Max-Planck-Institute for Physics of Complex Systems,\\
N\"othnitzer Strasse 38, D-01187 Dresden, Germany\\
Fax number: 0049-351-8711999\\
Department of Physics, University of Dschang P.O.Box 67 Dschang,
Cameroon\\}
\maketitle
\begin{abstract}
The bifurcation structure of coupled periodically driven double-well Duffing
oscillators is investigated as a function of the strength of the driving
force $f$ and its frequency $\Omega$.  We first examine the stability of
the steady state in linear response, and classify the different types of
bifurcation likely to occur in this model.
We then explore the complex behaviour associated with these bifurcations
numerically.  Our results show many striking departures
from the behaviour of coupled driven Duffing Oscillators with single
well-potentials, as characterised by Kozlowski et al~\cite{k1}.
In addition to the well known routes to chaos already
encountered in a one-dimensional Duffing oscillator, our model exhibits
imbricated period-doubling of both types, symmetry-breaking, sudden chaos
and a great abundance of Hopf bifurcations, many of which occur more
than once for a given driving frequency.
We explore the chaotic behaviour of our model using two indicators,
namely Lyapunov exponents and the power spectrum.
Poincar\'e cross-sections and phase portraits are also plotted to show
the manifestation of coexisting periodic and chaotic attractors
including the destruction of $T^2$ tori doubling.
\end{abstract}

\section{Introduction}
In recent years, a large number of theoretical calculations, numerical
simulations and experiments has been carried out on systems of coupled anharmonic oscillators
which provide fundamental models of the dynamical problems in several disciplines. Such
models describe a wide range of processes, helping in general to understand
the routes to chaos that take place in biological, chemical, physical,
electromechanical and electronical systems. Here we refer only to recent papers close
to the subject of our own[1-18]. Coupling may arise between oscillators of the same types or
oscillators of different types. Amongst these systems, the most intensively
investigated examples are the Duffing oscillators[1-8] (strictly dissipative
systems which converge to quiescence when not driven), the Van der Pol
oscillators[9-11] (self-excited systems which final behavior when not
driven is a limit cycle) and the coupled Van der Pol - Duffing oscillators [12-13].\\
In this paper we present the results of an investigation of two identical
coupled double-well Duffing oscillators subjected to a periodically driven
force. Our study was stimulated by the earlier work of Kozlowski et
al.~\cite{k1} who have studied the model of the same type, but considered only single-well
potentials. Using bifurcation diagrams and phase diagrams these authors
have demonstrated that the global pattern of bifurcation curves in
parameter space consists of repeated subpatterns similar to the superstructure
found in a one-dimensional driven Duffing oscillator. They have also
proven the existence of a Hopf Bifurcation which does not appear in a
model of one-dimensional Duffing oscillator. Other
interesting results have been reported in the context of coupled Duffing
Oscillators. In ref.~\cite{k2} Kunick et al showed that two
coupled oscillators can behave regularly with nonzero coupling and
chaotically with zero coupling. 
In a model of two coupled Duffing oscillators driven with incommensurate
frequencies and coupled additively, Stagliano et al~\cite{k3} observed the
period-doubling of an attracting $T^2$ torus and its destruction 
in parameter space. Migration control in two Duffing oscillators by
open-plus-closed-loop control method and adaptive control algorithm were studied
by Paul et al~\cite{k4}. These authors, in another system of two coupled
Duffing oscillators, have also investigated a basin of attraction with several
coexisting chaotic attractors and synchronization of chaos~\cite{k5}. Yagasaki~\cite{k6},
after having described an Auto driver Hommap for numerical analysis of
homoclinic in maps of periodically forced Duffing systems, modified a
version of the homoclinic Melnikov method for orbits homoclinic to two types of
periodic orbits. These theories have been successfully applied to a weakly
coupled Duffing oscillators. Moreover Yin et al~\cite{k7} investigated the
effect of phase difference in the driving forces of two coupled identical
Duffing oscillators. They observed desynchronization and lag
synchronization of chaotic attractors. Bifurcation behaviors, showing a kind of
Hopf bifurcation, from synchronous
chaos of a chain of Duffing oscillators have been explored by Ma Wen-Qi et
al.~\cite{k8} using the generalized  winding number in tangent space.\\
The equations of motion for the two identical coupled double-well Duffing
oscillators that we are interested in are the following:
\begin{eqnarray}
\label{e1}
\frac{d^2x}{dt^2} & = & -\alpha \frac{dx}{dt}+x-x^3+k(y-x)+f\cos (\Omega t)\\
\frac{d^2y}{dt^2} & = & -\alpha \frac{dy}{dt}+y-y^3-k(y-x)\nonumber
\end{eqnarray}
where $\alpha$ is the damping parameter, $k$ the coupling parameter, $f$ and
$\Omega $ the amplitude and the frequency of the driving force,
respectively. In system (\ref{e1}), we represent the oscillators by the state variables 
$(x,dx/dt)$ as oscillator $A$ or subsystem $A$ subjected to the periodic force
$f(t)=f\cos (\Omega t)$, which is coupled to the other one 
with the state variables $(y,dy/dt)$ as oscillator $B$ or subsystem $B$. Here
$f$ and $\Omega$ are the control parameters of our system. As evident, for
$k=0$, equation (1) describes two uncoupled Duffing Oscillators with their
motion governed by $\alpha$ and $f$. The coupling considered can be
interpreted as a perturbation of each oscillator through a signal proportional
to the difference of their amplitudes. A natural question to ask is how
chaotic attractors arise as these parameters are varied. In the case of 
one dimensional double-well Duffing oscillator, quasiperiodic
and period-doubling routes to chaos have been found~[19-27].
\\The rest of the paper is structured as follows. In section II, we establish
a linear stability analysis of the system projected on the Poincar\'e map. Section
III is devoted to the survey of bifurcation diagrams and the chaotic behavior
is characterized in section IV. We end with the conclusions of the work in
section V. 
\section{Stability analysis}
For stability analysis and even for purposes of numerical simulations,
it is convenient to transform the second order differential equation (1) into
an autonomous system of first order differential equations of the following
form:
\begin{eqnarray}
\label{e2}
\frac{dx}{dt} & = & v\nonumber\\
\frac{dv}{dt} & = & -\alpha v+x-x^3+k(y-x)+f\cos (\theta )\nonumber\\
\frac{dy}{dt} & = & w\\
\frac{dw}{dt} & = & -\alpha w+y-y^3-k(y-x)\nonumber\\
\frac{d\theta}{dt} & = & \frac{\Omega}{2\pi}\nonumber
\end{eqnarray}
or equivalently
\begin{eqnarray}
\frac{dV}{dt}=F(V,\mu) \nonumber
\end{eqnarray}
where $T={2\pi}/{\Omega}$ is the period of the driving force, $\theta$
the cyclic variable, $V(x,v,y,w,\theta)$ an autonomous vector field and
$\mu(\alpha,k,f,\Omega)$ an element of the parameter space. This system
generates a flow $\phi =\{ \phi^{T}\}$ on the phase space $\mathbf{R}^4 \times
{S}^1$and there exists a global map:
\begin{eqnarray}
\qquad \qquad P: \qquad \Sigma_{c}   \qquad \rightarrow \qquad   \Sigma_{c}\nonumber\\
V_{p}(x,v,y,w) \rightarrow P(V_{p})&=&\{ \phi^{T}\} |_{\Sigma_{c} (x,v,y,w,\theta_{0})}\nonumber
\end{eqnarray}
with $\theta_{0}$ being a constant determining the location of the Poincar\'e
cross-section and $(x,v,y,w)$ the coordinates of the
attractors in the Poincar\'e cross-section $\sum_{c}$ defined by: $\sum_{c}=\{(x,v,y,w,\theta) \in \mathbf{R}^4 \times {S}^{1} |
\theta=\theta_{0}\}$. System (\ref{e2}) is symmetric since the transformation 
$S:(x,v,y,w,\theta) \rightarrow (-x,-v,-y,-w,\theta+ \pi)$ leaves (\ref{e2})
invariant. This system can support symmetric orbits and also asymmetric ones which
are not invariant with the transformation S. We use a perturbation
analysis to proceed analysis of solution stability in the poincar\'e map $\Sigma_{c}$. This method consists of differentiation and translation. Thus if we
add a small perturbation $\delta V_{p}$($\delta x$, $\delta v$, $\delta y$,
$\delta w$) to the steady state $V_{p0}$($x_{0}, v_{0}, y_{0},w_{0}$) we obtain equations for perturbed motion
\begin{equation}
\label{e3}
x=x_{0}+\delta x,  y=y_{0}+ \delta y, 
v=v_{0}+\delta v,  w=w_{0}+ \delta w
\end{equation}
Substituting equations (\ref{e3}) into equations (\ref{e2}) and developing into powers of the
above small perturbations yields the following variational differential
equation:
\begin{equation}
\label{e4}
\frac{d{\delta V_{p}}}{dt}=DF(V_{p0})\delta V_{p}, 
\end{equation}
with \qquad
\[
DF(V_{p0})= \left( \begin{array}{cccc}
0 & 1 & 0 & 0 \\
1-k-3x_{0}^2 & -\alpha & k & 0\\
0 & 0 & 0 & 1\\
k & 0 & 1-k-3y_{0}^2 & -\alpha\\
\end{array} \right) \]
where $DF(V_{p0})$ is the Jacobian $4 \times 4$ matrix describing the vector field along
the solution $\delta V_{p}(t)$ and $V_{p0}$ being an equilibrium
point or the steady state. Next for the very small amplitudes, we neglect in
the matrix  $DF(V_{p0})$ the powers higher than one, that is, $x_{0}^2$ and $y_{0}^2$. The solution after one period $T$ of
the oscillations in the linearised Poincar\'e map is simply expressed as:
\begin{equation}
\label{e5}
\delta V_{p}(T) = \delta V_{p0} \exp (DF(V_{p0})T),  \qquad \delta V_{p}(0)=\delta V_{p0} 
\end{equation}
where $DF(V_{p0})$ is the monodromy matrix of a periodic orbit connecting arbitrary
infinitesimal variations in the initial conditions $\delta V_{p0}$ with
corresponding changes $\delta V_{p}(T)$, after one period T. The stability of
the periodic motion is therefore determined according to the real parts of the
roots of the characteristic equation $det(DF(V_{p0})-{I}\lambda)=0$ written in
the following form:
\begin{equation}
\label{e6}
\lambda^4+A_{3}\lambda^3+A_{2}\lambda^2+A_{1}\lambda+A_{0}=0
\end{equation}
where $(A_{i})$,$i=0,1,2,3$ are the coefficients depending on the two
parameters $k$, $\alpha$ and $\lambda$ =$(\lambda_i)$ the eigenvalues of $DF(V_{p0})$. The roots of equation
(\ref{e6}) are obtained using the Bairstow-Newton-Raphson algorithm~\cite{k28} and for different
values of $k \in [-5,5]$ and $\alpha \in [-0.2,0.2]$. The results as the
spectrum of the eigenvalues are depicted in a
complex plane $\mathbf {C}$ (see fig.1). This figure enables us to get an idea
of different types of bifurcation likely to appear in the system. Since $DF(V_{p0})$
is a real matrix, complex eigenvalues occur in a complex conjugate pairs
responsible of the symmetry observed along the real axis. Thus if $\lambda_{i}$
is real, it is clear from equation (\ref{e5}) that the eigenvalues is the rate of
contraction ($\lambda_{i}<0$) or expansion ($\lambda_{i}>0$) near the steady
state. Next if $\lambda_{i}$ is complex, the real part of $\lambda_{i}$ gives
the rate of contraction $(Re(\lambda_{i})<0)$ or expansion
$(Re(\lambda_{i})>0)$ of the spiral while the imaginary part
$Im(\lambda_{i})$ is of the frequency rotation. This can be well understood
with the following expression of the eigenvalues of the linearized Poincar\'e
map:
\begin{equation}
\sigma_{i}=exp(Re(\lambda_{i}))[cos(Im(\lambda_{i})T)+jsin(Im(\lambda_{i})T)]
\end{equation}
Globally, if $Re(\lambda_{i})<0$ for all $\lambda_{i}$, then all sufficiently
small perturbations tend toward zero as t goes to infinity and the steady state
(nodes (n), saddle nodes (sn), spiral (sp)) is stable. If $Re(\lambda_{i})>0$ for all
$\lambda_{i}$, then any small perturbation grows with the time and the steady
state $(n, sn, sp)$ is unstable (A stable or unstable equilibrium state with no
complex eigenvalues is often called node). On the other hand if there exists
$i$ and $l$ such that $Re(\lambda_{i})<0$ and $Re(\lambda_{l})>0$, therefore the
equilibrium state is non-stable (A non-stable equilibrium state is often
called saddle; an equilibrium point whose eigenvalues all have a non zero real
part are called hyperbolic~\cite{k29}).
Having regards to the precedent with again a look on the spectrum of fig.1,
it follows that our system of the given parameter {k} and $\alpha$ can undergo many types
of bifurcations namely: saddle-node $(Re(\lambda_{i})=1)$, period-doubling
$(\lambda_{i}=-1)$, Hopf bifurcations ($\lambda_{i}=\gamma \pm j \beta~, (j^{2}=-1$) with
$\gamma <0$ and the motion has the form of exponential increasing beats with
period ${2\pi}/{\beta}$~\cite{k17}) and of course symmetry-breaking
bifurcation which is often a prerequisite for the first period-doubling
bifurcation~\cite{k30}. Except Hopf bifurcation, such bifurcations  have been
successfully found in one-dimensional double-well Duffing oscillators
subjected to a periodically driven force~[19,21,26] using the eigenvalues called Floquet
multipliers of the linearized Poincar\'e map. In a coupled identical single-well
Duffing oscillators, Kozlowski et al~\cite{k1} showed by linear analysis a similar eigenvalues scenario and they conjectured that the
resulting alternating bifurcation sequences have to be expected for all
systems of coupled strictly dissipative oscillators that are driven periodically.
The above analysis allows us to know how the steady state becomes locally
unstable and to be aware of the type of bifurcation expected in the
system. Here, no exploding amplitude is possible since the shape of the potential offers globally bounded solutions $(V(x,y)=-(1/2)x^2-(1/2)y^2) +(1/4)x^4+(1/4)y^4) \rightarrow +\infty \quad as \quad |x|,|y| \rightarrow \infty)$.

\section{Local Bifurcations}
The aim of this section is to seek numerically the routes which lead to
chaotic solutions of the system when the control parameter $\Omega$ evolves,
for different values of the driving force $f$. Bifurcation diagrams are very
good for numerical as well as for experimental studies when there is a tunable
parameter. When a control parameter is
varied and bifurcation takes place a qualitative change of the system
happens. For the numerical computations of these diagrams the control
parameter $\Omega$ is increased from an initial value $\Omega_{i}$ to a final value
$\Omega_{f}$ and then decreased from $\Omega_{f}$ to $\Omega_{i}$ in a very small
step ($\Delta \Omega=10^{-5}$). The last computed cyclic point for a given
value of $\Omega$ is always used as a new initial value for the next value of
$\Omega$. Starting with initial conditions $(x,v)=(1,0), (y,w)=(1,0)$ at
$\Omega_{i}$, system (\ref{e2}) is integrated, using the standard Runge Kutta
Algorithm~\cite{k28},  for $100$ periods of the driving force until the transient
has died out; the trajectories expected are attractors and the local
calculation error is sufficiently small. Then to find out whether the
trajectory is periodic (quasiperiodic) or chaotic, the system is integrated
for the next $180$ periods in order to catch the maximum of
coexisting attractors. This procedure allows us to make sure that coexisting
attractors are realised due to  different initial conditions which are handed
throughout from one parameter to another. We chose $k=5$ and $\alpha=0.1$ for
all examples in the subsequent section. The bifurcation diagrams obtained
here show the projection of the attractors in the Poincar\'e section onto $x$
or $v$ versus the control parameter $\Omega$ and for different values of
$f$. In a one-dimensional double-well Duffing oscillator~[19-21,23-27], it is shown
that chaotic solutions arise through quasiperiodic, period-doubling cascade
with a sequence of symmetry-breaking and saddle-nodes. Hopf bifurcations have
not been found, but are abundant in our model.\\
We computed a great number of diagrams, and some of them showing
the main features of the system are presented in figures 2-4. It is important
to note that no hysteresis has been detected while checking all these
diagrams. We can therefore avoid any confusion over
the periods of attractors when reading this type of diagram; for instance
one single period-2 yields to 2 points for one frequency value whereas two
coexisting attractors each of period-1 also yield to two points for one
frequency. These diagrams show a great number of coexisting attractors
(chaotic domains) intermingled with imbricated windows made up of periodic
solutions of different periodicity, period-doubling of both types, sudden
Chaos and Hopf bifurcation. The so called chaotic domains (or chaotic sea) will be characterized in the following section.\\
At lowest frequencies ($0<\Omega \leq 0.4$) quasiperiodic as well as chaotic
solutions occur together with a number of symmetry-breaking (sb), period
doubling (pd) and saddle-nodes (sn). In fig.2(a) where $f=15$, it
appears in the neighborhood of $\Omega =0.385$, 6 pairs of sb followed by
3 reversed pd till the chaotic domain. When the driving force is increased
$f=20$, overlapping pd cascade and sn appear as can be seen in fig.2(b).
Similar features as in the latter case are also observed for small values of
$f$ ($f<15$). In other ranges of the driving frequency bigger than $\Omega =0.4$ the
aformentioned features still exist and additional ones such as resonances and
Hopf bifurcations appear in the system; the results are displayed in figures 3
and 4. In figure 3 it can be seen throughout overlapping pd cascade of both types, sb, resonances (R) and even sudden chaos (SC). When $f=4.5$
in fig.3(f) at $\Omega \simeq 0.785$, period-8 attractor undergoes reversed
period-doubling cascade and period-2 is created as from $\Omega \simeq 0.805$. Besides, for
$f\ge 7$ appear resonances (R) in fig.3(a,b,d,e), fig.6(a) and sudden chaos
in fig.3(c).
However in fig.4 when $f$ becomes much larger and $\Omega \ge 0.5$, Hopf Bifurcation
(H) is abundant and often appear more than once at a given frequency. This
notation $-nH-$ means that there are $n$, ($n\leq 3$), integer Hopf bifurcation at a given
frequency. One can therefore observe -H- in (a), -3H- in (b), -2H-..-2H- in (c,d),
-2H-..-3H-..-2H- in (e,f) and -3H-3H- in (g,h). In fig.4(h) for instance, at
$\Omega=0.5925$, a period-6 attractor undergoes
reversed period-doubling to become a period-3; next after 3 pairs of
symmetry-breaking near $\Omega \simeq 0.6$, appears again period-3 attractor as from
$\Omega \simeq 0.601$. And then at
$\Omega =0.603$ this attractor of period-3 undergoes 3 Hopf bifurcations. Windows of periodic
solutions separated by quasiperiodic ones are clearly visible. The Poincar\'e
cross-section of quasiperiodic attractors of this kind (-3H-) consists of three invariant tori
(destroyed or not). Thus at any $\Omega $ where $-nH-$ exist, the
projection of the attractor corresponds to $n$~Tori on the
Poincar\'e cross-section. The torus attractor arises from Hopf bifurcation
which is a bifurcation from a fixed point to an invariant curve. At the transition to
quasiperiodic motion, this fixed point (projection of the limit cycle in to
the cross-section) loses its stability and gives birth to an invariant circle,
which is a cross-section of the torus in the flow.  The three pairs of sb appearing at the neigborhood of
$\Omega \simeq 0.6$ (fig.4(g,h)) seem to be apparently a window of period-6 attractor. In
fact it is just a period-3 attractor as can be illustrated further in
fig.8(b).This orbit, born asymmetrically, coexists with its inversion
symmetric orbit of the same period-3 and are caught simultaneously.\\
Next we pursue our investigation for large values of $\Omega $ ($\Omega \ge 3$)
and for different values of $f$. It turns out that the features observed so far have disappeared and the
system becomes regular providing attractors with a small number of periods as can be seen in
figure 5. The phase diagrams which often show bifurcations, curves, surfaces
and which is out of the scope of this paper could also help to match the
features obtained in this analysis.
 
\section{Chaotic behavior}
To better support the results obtained above, we intend in
this section to characterize chaotic behavior. For this purpose some
indicators  are used namely the Lyapunov exponents and the power spectrum
density. Also, Poincar\'e cross-sections and phase portraits are shown to
illustrate some attractors present in the system. 
The Lyapunov exponent is one of the most important tools for understanding chaotic
behavior, obtained by examining ``a very sensitive dependence of initial
conditions". In particular it is generally well established that a rigorous
measure of chaos may be given in terms of Lyapunov spectrum of the dynamical
system. A positive Lyapunov exponent is characteric of chaos while zero and
negative values of the exponent signify a marginally stable or quasiperiodic
orbit and periodic orbit, respectively. Solving numerically the variational
equation (\ref{e4}) together with system (\ref{e2}) we calculate the maximum
Lyapunov exponent $\lambda_{max}$ in the Poincare cross-section  with
$\lambda_{max}=\lim_{\tau \rightarrow +\infty} \frac{1}{\tau} \ln (\|L(\tau)\|)$,
where $\|L(\tau)\|= \Big(\delta x^2+\delta v^2+\delta y^2 +\delta
w^2\Big)^{1/2}$. The spectrum is displayed in fig.6 together
with a bifurcation diagram. Windows of chaotic domains and periodic solutions are clearly justified.
Power Spectrum density using Fast Fourier Transformation (FFT) has also been
computed with some values from fig.6 for a periodic solution ($\Omega=0.495$,
 $f=7$) and a chaotic solution ($\Omega =0.47$,$f=7$) (see fig.7).
Fig.8 shows the phase portraits and Poincar\'e cross-section. In fig.8(a,b)
plotted are the phase portraits, related to fig.4(g,h), showing a period-3 attractor for $f=30$, $\Omega=0.6025$(a) and a coexistence of two asymmetric attractors, each of period-3 for
$f=30$, $\Omega = 0.6$ (b). The dots ($\bullet$) represent the corresponding
points in the Poincar\'e cross-sections. In fig.8(c,d) two chaotic attractors
are plotted in the Poincar\'e cross-section for $\Omega =0.5$, $f=2$ and
for $\Omega = 0.375$, $f=15$, respectively. Furthermore, we observe in
fig.8(e,f) three $T^2$ tori doubling at $\Omega= 0.605$, $f=30$ and at
$\Omega =1.14$, $f=15$, two $T^2$ tori doubling (destroyed) corresponding for
-3H- and -2H-, respectively (values from fig.4(c,h)). In the torus route to chaos, the original torus
appears to split into two circles at the torus doubling bifurcation
point. This route is reminiscent to the period-doubling route to chaos,
although there are finite number of torus doublings before the onset of chaos
motion. Also, the torus doubling route to chaos is a higher-dimensional
phenomenon,  requiring at least a four-dimensional flow, or a three-dimensional
map. It is not observed in one-dimensional maps, unlike the period-doubling
route to chaos~\cite{k30}.

\section{Conclusion}
To sum up we have investigated coupled double-well Duffing oscillators
subjected to a periodic driving force. Our main conclusion is that the shape
of the potential in the coupled Duffing Oscillators system has a profound
influence on the routes to chaos taken by the system. By a linear
analysis we have examined the stability of the steady state of the system
leading to different types of bifurcation likely
to appear for $k \in [-5,5]$ and $\alpha \in [-0.2,0.2]$. With the help of bifurcation diagrams, it has been shown that the
bifurcation structure, which is complicated, depends strongly on the values
of the control parameter $\Omega$. For a simple electronic realization of our model, it is obvious that
the oscillators may be very difficult to control for small values
of $\Omega$ due to the large windows for chaos. This system becomes regular for large
values of $\Omega$ ($\Omega \geq 3$). Apart from the routes already encountered
in a one-dimensional double-well Duffing oscillator such as quasiperiodic and
period-doubling cascade, sudden chaos and mostly Hopf bifurcations have been
newly found in this work. Besides, resonances (R), imbricated sb, sn, pd
intermingled by chaotic domains are observed, rending the structure
highly chaotic. Two and three $T^2$~tori doubling, have been also observed. To characterize chaotic
behavior of this system, Lyapunov
exponents and the power spectrum density using FFT have been employed. Poincar\'e
cross-sections and phase portraits have been also used for better
visualisation of periodic and chaotic attractors.
An analytical approach using approximative technics could be an interesting
topic for future work, and would probably shed further light on the rich
behaviour of this model.\\

\section{ACKNOWLEDGMENTS}
It is a pleasure to thank Prof. Dr. Jan M. Rost for illuminating discussions
and Dr. Nic Shannon for a careful reading of the manuscript. Financial support
by Alexander von Humboldt (AvH) Foundation/Bonn-Germany,
 under the grant of Research fellowship No.IV.4-KAM 1068533 STP is gratefully acknowledged.

\newpage
\section{{\bf Figure caption}}
\begin{itemize}
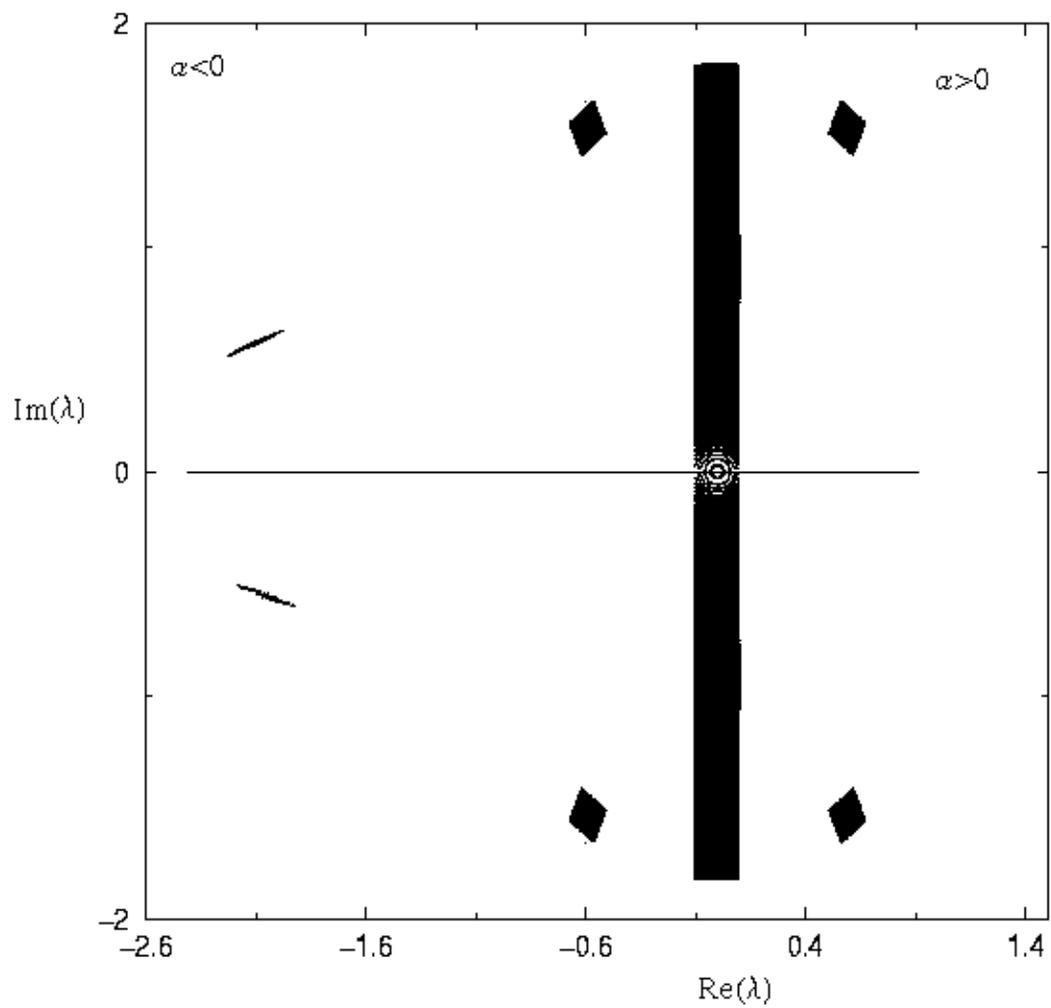
\item {Figure 1: 
Eigenvalues spectrum for the stability of the system with $\alpha \in
[-0.2,0.2]$ and $k \in [-5,5]$. Note that when $\alpha <0$, the eigenvalues are
in the half-plane left and in the half-plane right otherwise.}

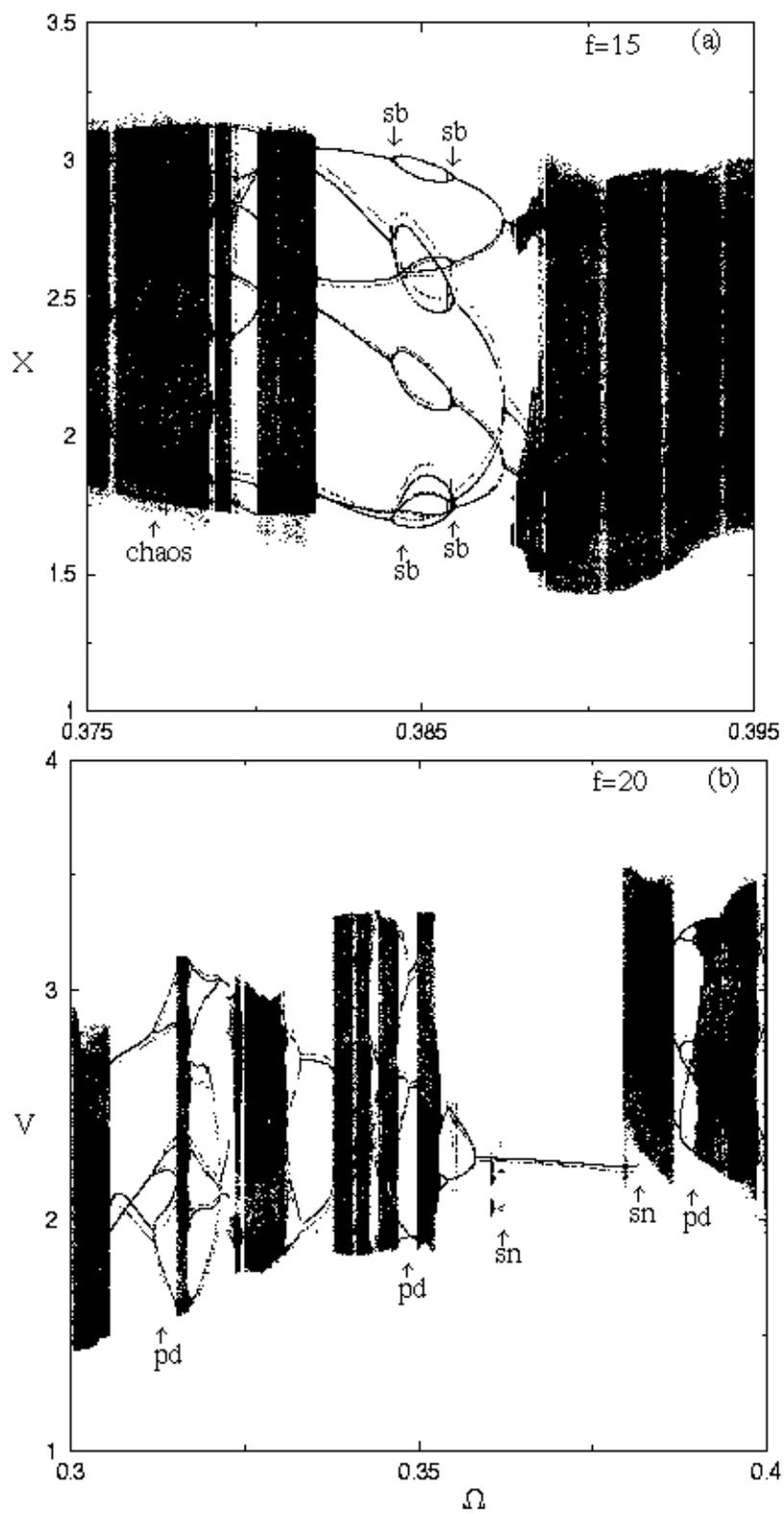
\item {Figure 2: 
Bifurcation diagrams showing the first coordinate $x$ for the driving amplitude
$f=15$~$(a)$ and the second coordinate $v$ for $f=20$~$(b)$ of the
Poincar\'e section versus the driving frequency $\Omega$. Windows of periodic
solutions and chaotic domains are visible. Near $\Omega = 0.385$~(a), $6$
pairs of $sb$ occur followed by the reversed $pd$ till
the chaotic domain. Imbricated $pd$, $sb$, $sn$ are mentioned. Note that
similar behavior as (b) is observed when $\Omega \leq 0.4$ and for small values
of $f$ ($f<15$).}

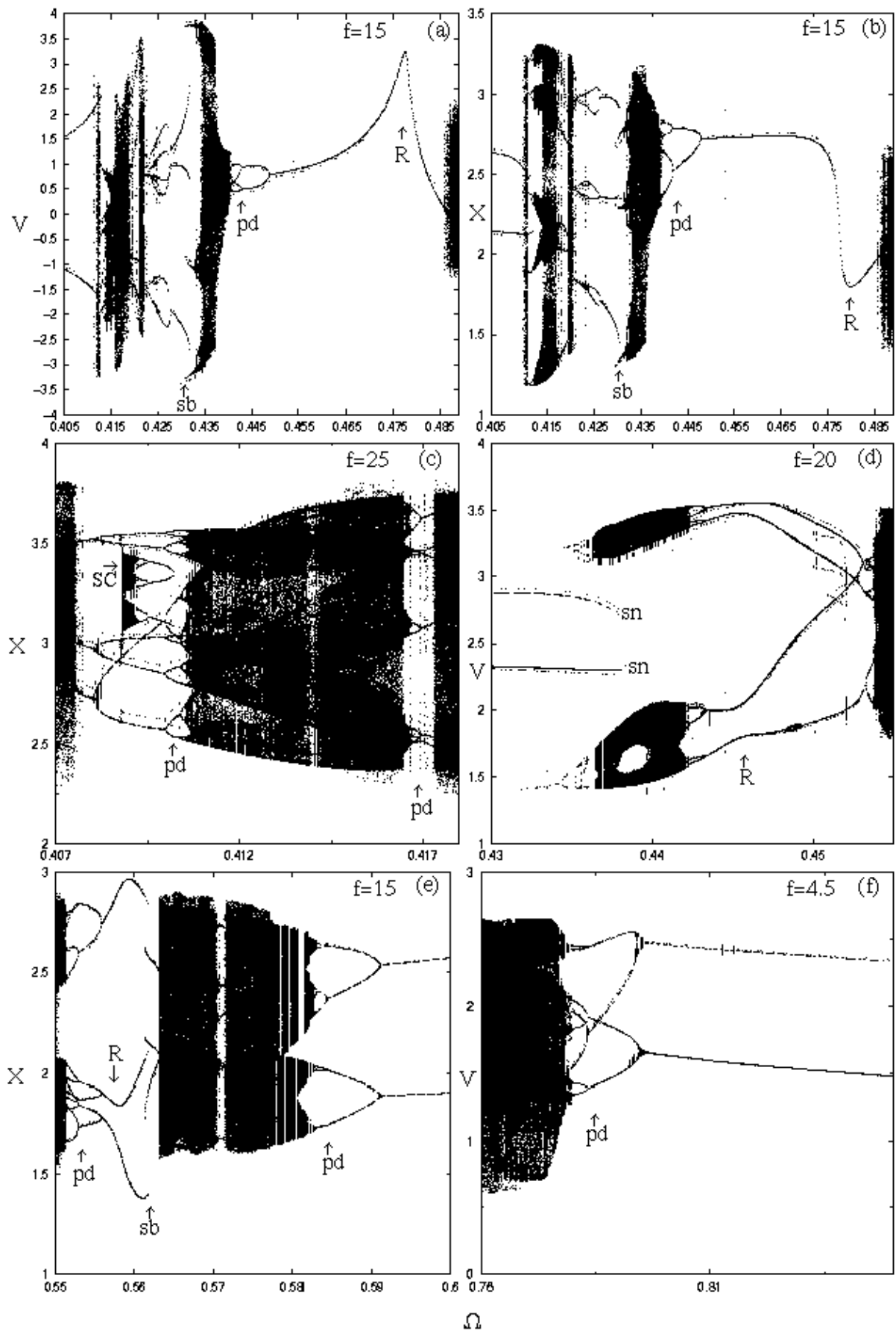
\item {Figure 3: 
Bifurcation diagrams for the driving force $f=4.5$~(f), $f=15$~(a,b,e),
$f=20$~(d), $f=25$~(c) with $\Omega > 0.4$. Throughout imbricated $sb$, $sn$,
$pd$ of both types and chaotic domains are observed.  Besides, resonances (R)
appear at $\Omega \simeq 0.475$~(a,b), $\Omega \simeq 0.445$~(d), $\Omega
\simeq 0.555$~(e)and sudden chaos at $\Omega \simeq 0.408$ (c).}

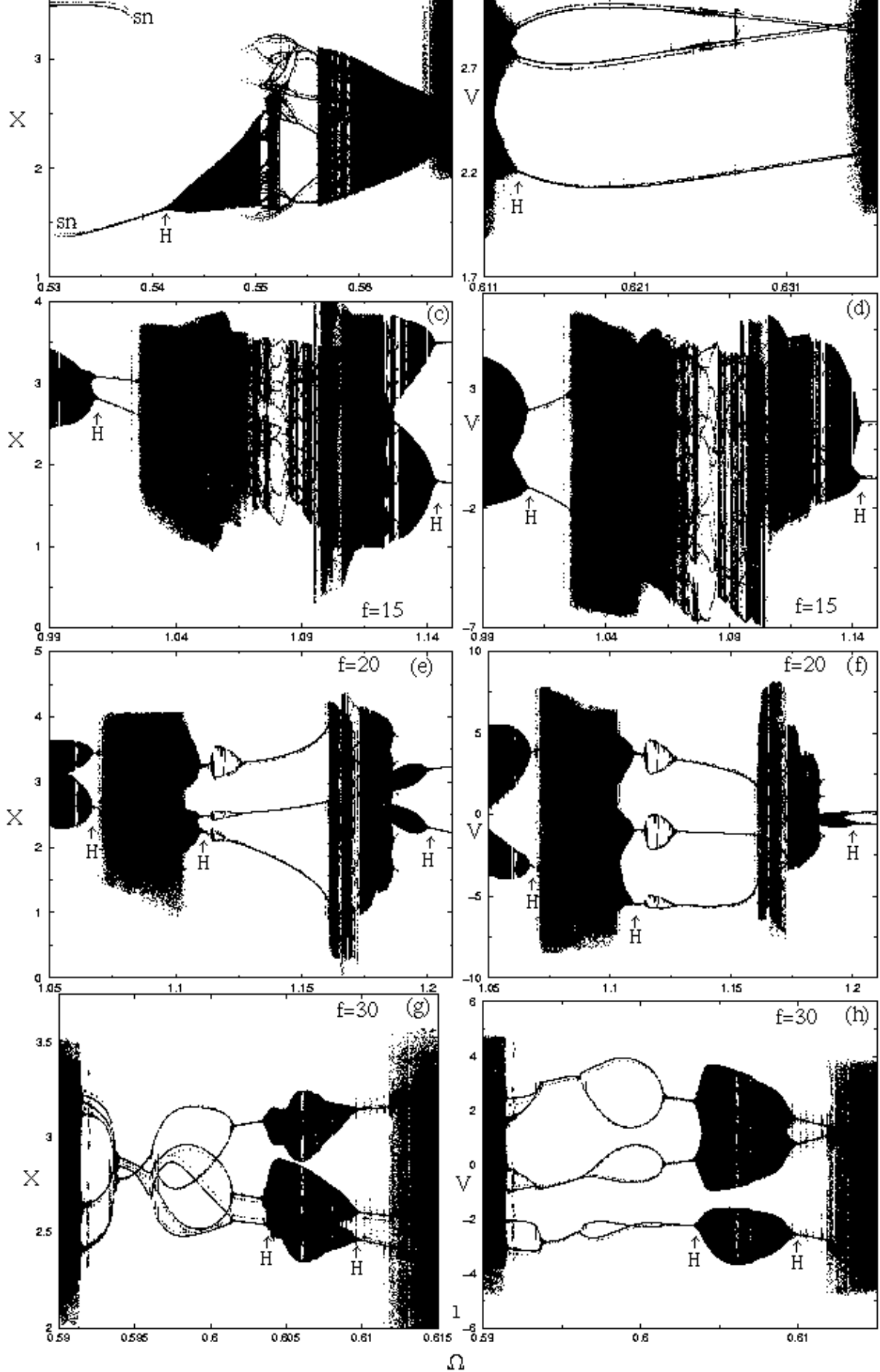
\item {Figure 4:
Bifurcation diagrams for $f=30$~(a,g,h), $f=25$~(b), $f=15$~(c,d),
$f=20$~(e,f) showing many Hopf bifurcations (H). One can observe -H-(a), -3H-(b), -2H-..-2H (c,d),-2H-..-3H-..-2H-(e,f), -3H-3H- (g,h) 
(see also text)}

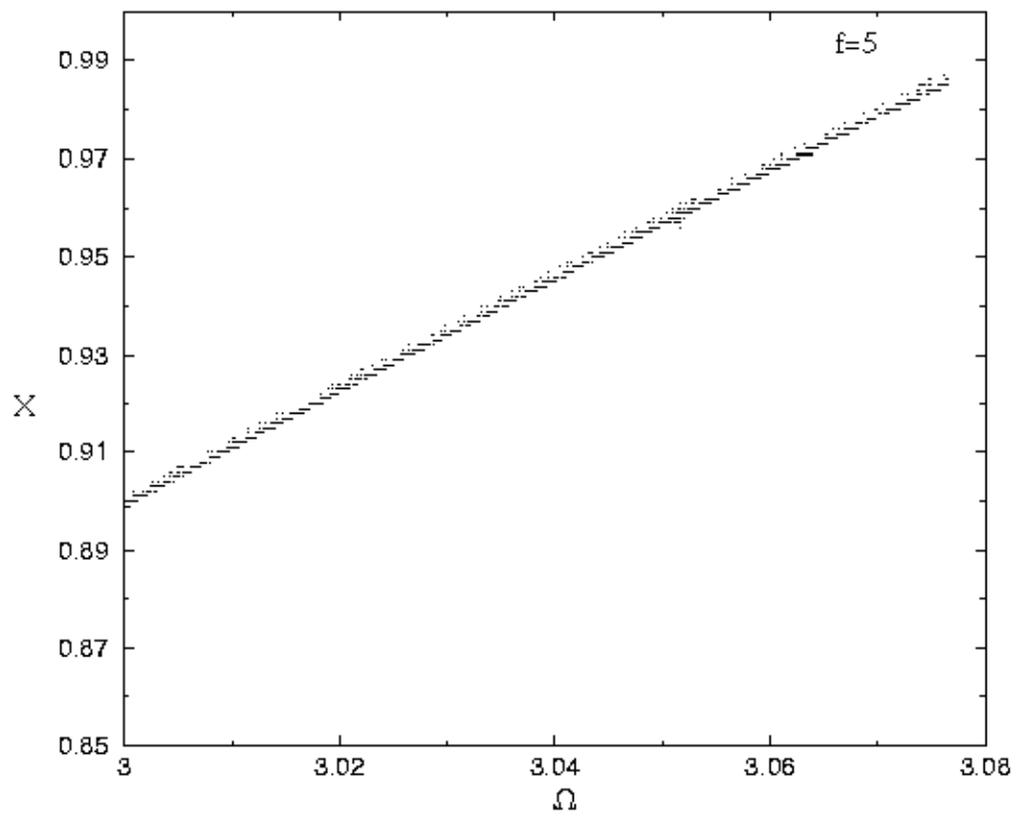
\item {Figure 5: 
Bifurcation diagram for $f=5$ showing attractors of small number of
periods. Note that the system becomes regular and this behavior is observed
for $\Omega \geq 3$ and for any value of $f$.}

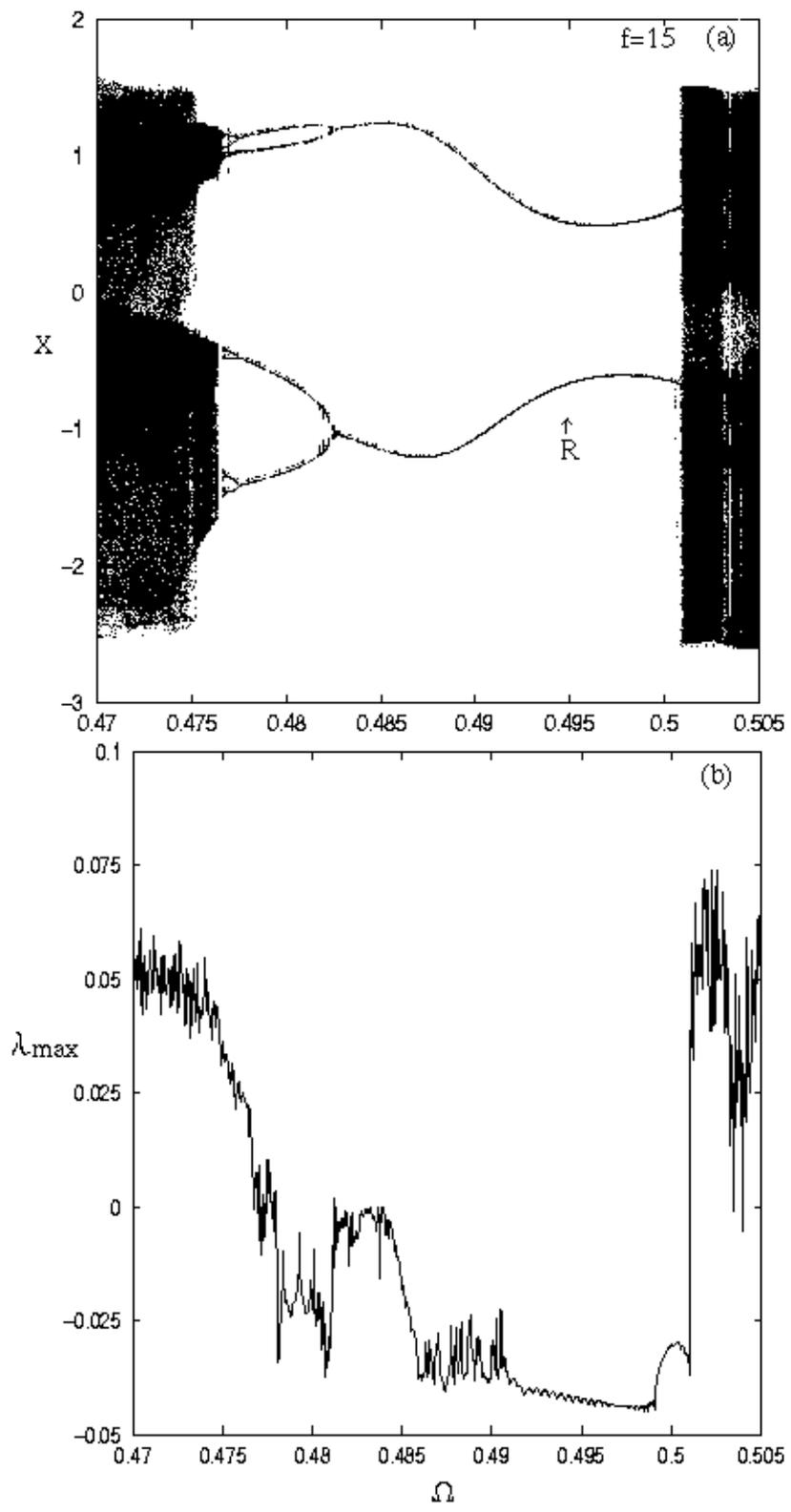
\item {Figure 6:
Bifurcation diagram for $f=15$~(a) and the corresponding Lyapunov
spectrum (b) in the Poincar\'e map. windows of chaotic domains are clearly
justified with positive values of $\lambda_{max}$.}

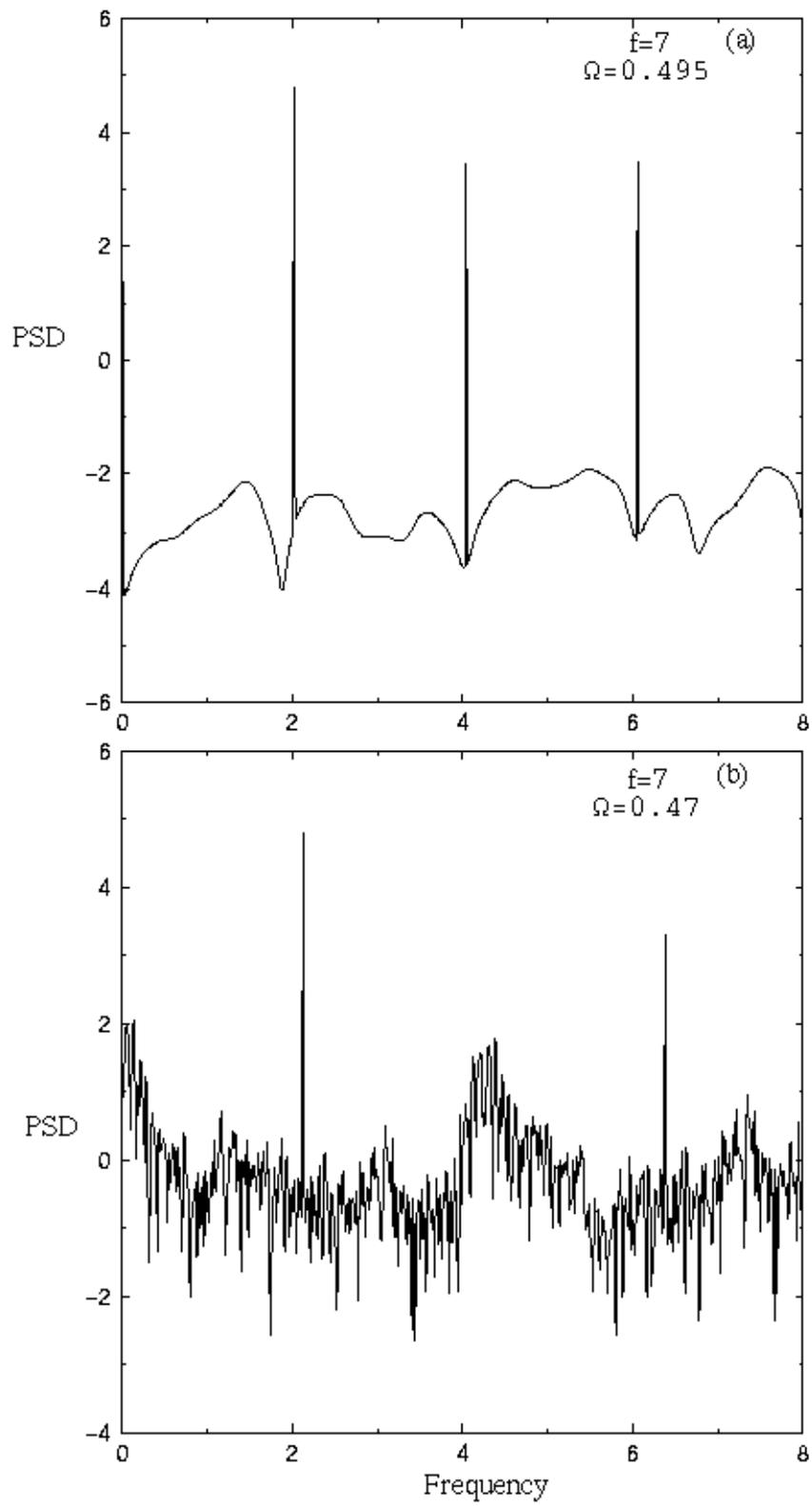
\item {Figure 7:
Power spectra densities (PSD). The values of $f=7$ and $\Omega$ are from the
bifurcation diagrams shown in fig.6 with $\Omega = 0.495$~(a) periodic motion and $\Omega =0.47$~(b) chaotic motion.}

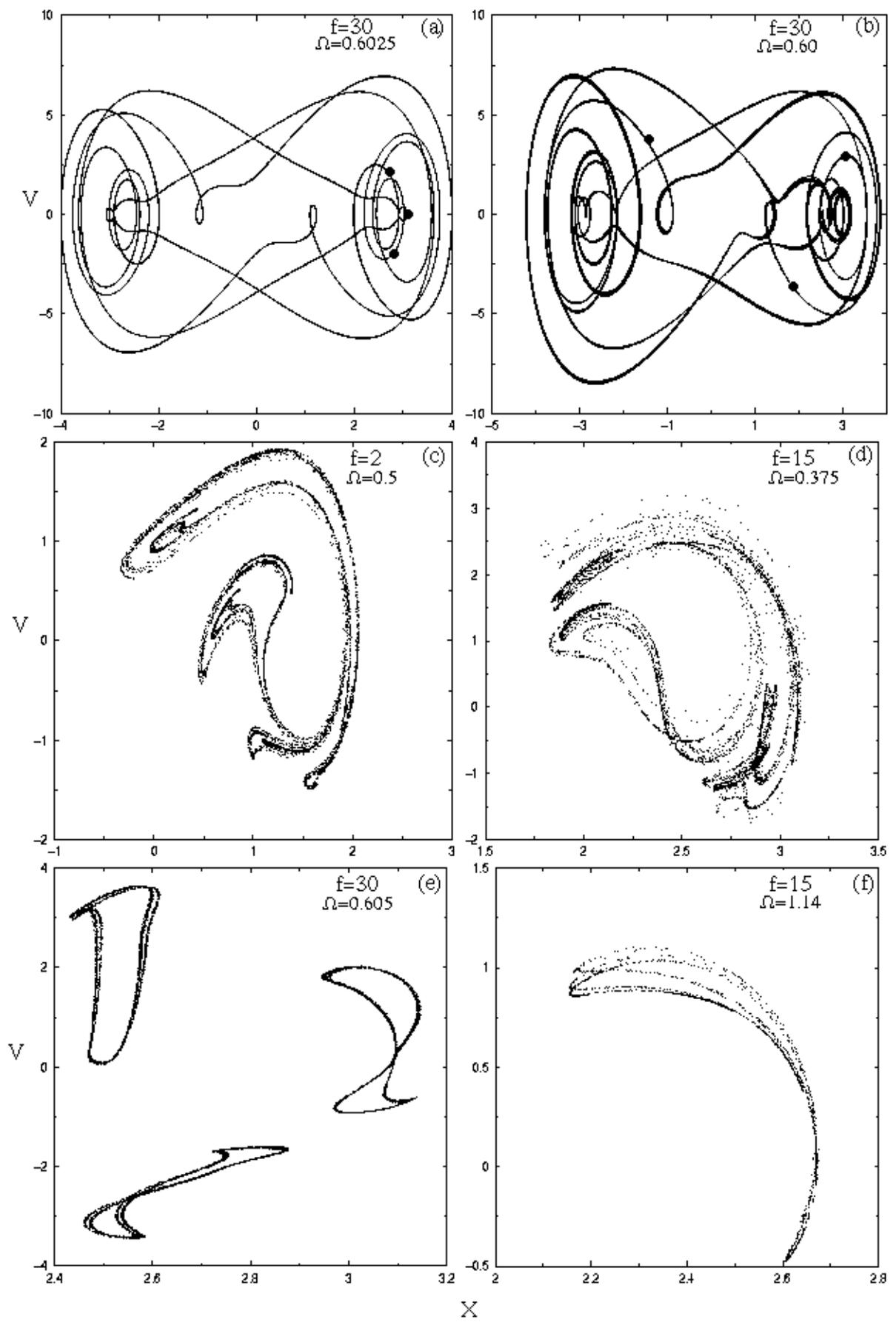
\item {Figure 8 :
Phase portraits and projection of the attractors in the Poincar\'e section onto
the two first coordinates $x,v$ of the Poincar\'e cross-section. At $\Omega =
0.6025$~(a), one period-3 orbit and at $\Omega = 0.6$ (b) coexistence of two
asymmetric orbits of the same period-3. The values of $\Omega $ are
from fig.4~(g,h) and $f=30$. The dots ($\bullet$) represent the corresponding points in the
Poincar\'e cross-section. Two chaotic attractors shown in (c,d) for
$\Omega=0.5$, $f=2$ and $\Omega = 0.375$, $f=15$, respectively. Three $T^2$
tori doubling for $\Omega = 0.605$,
$f=30$ (e) and two $T^2$ tori doubling (destroyed) for $\Omega = 1.14$, $f=15$
(f). The values of $\Omega $ are from fig.4(c,h).}
\end{itemize}
\end{document}